# Graphene-Flakes Printed Wideband Elliptical Dipole Antenna for Low Cost Wireless Communications Applications

Antti Lamminen, Kirill Arapov, Gijsbertus de With, Samiul Haque, Henrik G. O. Sandberg, Heiner Friedrich, and Vladimir Ermolov

*Abstract*—This letter presents the design, manufacturing and operational performance of a graphene-flakes based screen-printed wideband elliptical dipole antenna operating from 2 GHz up to 5 GHz for low cost wireless communications applications. To investigate radio frequency (RF) conductivity of the printed graphene, a coplanar waveguide (CPW) test structure was designed, fabricated and tested in the frequency range from 1 GHz to 20 GHz. Antenna and CPW were screen-printed on Kapton substrates using a graphene paste formulated with a graphene to binder ratio of 1:2. A combination of thermal treatment and subsequent compression rolling is utilized to further decrease the sheet resistance for printed graphene structures, ultimately reaching 4 Ω/□ at 10 μm thicknesses. For the graphene-flakes printed antenna an antenna efficiency of 60% is obtained. The measured maximum antenna gain is 2.3 dBi at 4.8 GHz. Thus the graphene-flakes printed antenna adds a total loss of only 3.1 dB to an RF link when compared to the same structure screen-printed for reference with a commercial silver ink. This shows that the electrical performance of screen-printed graphene flakes, which also does not degrade after repeated bending, is suitable for realizing low-cost wearable RF wireless communication devices.

## I. Introduction

PRINTED flexible graphene antennas for communications systems are at the focus of science and technology on account of their decent electrical, but more importantly, excellent mechanical properties. Recently, graphene-based RFID antennas [1], [2], RF transmission lines and antennas [3], and fully integrated RFID devices [4]–[6] have been demonstrated. Conductivity of printed graphene flakes is significantly lower than that of copper, aluminium and even that of printed metallic inks, which are the most commonly used conductor materials in flexible antennas to date. Nevertheless, graphene-based structures have several advantages such as low cost, chemical stability, mechanical flexibility and fatigue resistance. For comparison, the cost of silver ink depends strongly on the price level of bulk silver metal. The price of silver is very stable and is not expected to decrease in the foreseeable future. The cost for graphene ink raw materials is very low and the manufacturing process is simple and scalable. The cost can be estimated based on the complexity and scalability of the process and is expected to be significantly lower than commercially available silver inks.

The above mentioned planar dipole or slot-type antennas which have been realized [1]–[6] are narrow band and can be used only in a limited frequency range. There is also a strong demand for low cost wideband antennas. Furthermore, for wideband applications such as ultra-wideband (UWB) wireless body area networks (WBAN) [7]–[9], flexibility and bending fatigue resistance for antennas is required. Recently, a wideband graphene-printed triangular slot antenna design based on multiple resonances has been published in [10]. A downside of such antennas is usually multiple reflections in the antenna structure that may distort the UWB pulse in the time domain.

We present in this letter a graphene-flakes printed non-resonant planar elliptical element dipole antenna with a reasonably high operational efficiency, designed by taking into consideration the moderate conductivity of printed graphene flakes. We also examine the electrical properties of a graphene-flakes printed transmission line up to 20 GHz. Finally, we demonstrate the flexibility of a graphene-flakes printed sheet by showing a stable resistance for the printed sheet even after numerous bending cycles.

## II. Antenna and Transmission Line Design and Fabrication

An important consideration for the antenna design is that screen-printed graphene-flake conductors have a relatively

This work was supported by the EC under the Graphene Flagship under grant agreement no. 604391.
A. Lamminen, H. G. O. Sandberg, and V. Ermolov are with VTT Technical Research Centre of Finland, Tietotie 3, FI-02044, Espoo, Finland (e-mail: antti.lamminen@vtt.fi).
K. Arapov was with the Laboratory of Materials and Interface Chemistry, Department of Chemical Engineering and Chemistry, Eindhoven University of Technology, Eindhoven, De Zaale, 5612AJ, The Netherlands. He is now with Johnson Matthey Advanced Glass Technologies BV, Fregatweg 38 Maastricht, Limburg 6222 NZ, The Netherlands.
G. de With and H. Friedrich are with the Laboratory of Materials and Interface Chemistry, Department of Chemical Engineering and Chemistry, Eindhoven University of Technology, Eindhoven, De Zaale, 5612AJ, The Netherlands.
S. Haque was with Nokia R&D UK, 21 JJ Thomson Avenue, Madingley Road, Cambridge, CB3 0FA, The United Kingdom. He is now with Emberion Limited, Sheraton House, Castle Park, Cambridge CB3 0AX, The United Kingdom.

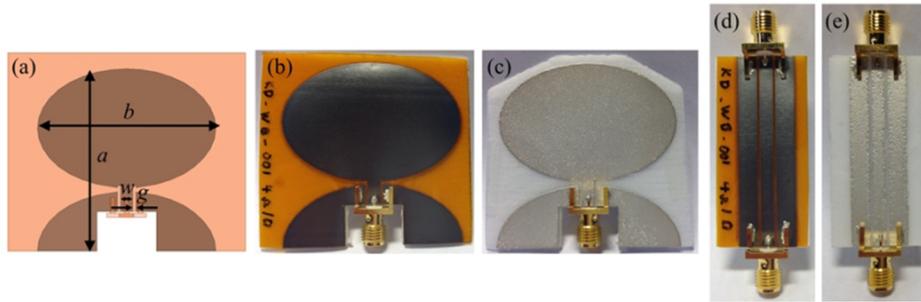

Fig. 1. Wideband quasi dipole antennas (WBQD) and coplanar waveguide (CPW) transmission lines. (a) EM simulation model of the WBQD antenna ($a$ = 46 mm, $b$ = 45 mm, $w$ = 4 mm, $g$ = 0.9 mm), (b) printed graphene WBQD antenna, (c) printed silver WBQD antenna, (d) printed graphene CPW, and (e) printed silver CPW.

high sheet resistance. Hence, we chose a wideband dipole antenna with elliptical branches [11], where the impact of loss resistance is minimized by maximizing the width-to-length ratio. Thus, antenna radiation efficiency ($\eta_r$) is maximised by keeping the radiation resistance ($R_r$) high and the loss resistance ($R_l$) low, following the basics of the antenna theory equation $\eta_r = R_r/(R_r+R_l)$ [12]. The antenna has a tapered impedance transformer from feed line to air and thus is not of a resonant type which is usually advantageous in wideband applications. The wideband quasi dipole antenna (WBQD) employs an axial ratio of 1.5 for the ellipse as this has been found to be an optimal value with respect to input matching and uniformity of the radiation pattern [11]. The final WBQD antenna layout (Fig. 1(a)) was designed using a commercial electromagnetic (EM) simulator (Ansys HFSS 2014) to operate at frequencies from 2 GHz up to 5 GHz. In the simulations a target DC sheet resistance of $R_s$ = 4 Ω/□ for graphene-flakes printed conductors was used. The WBQD is fed by a coplanar waveguide (CPW) transmission line shown in Fig. 1(d) and (e). Due to the fact that printed graphene transmission lines have a rather high attenuation at 2–8 GHz [3], the CPW length was minimised by halving one branch of the antenna dipoles. In addition, the antenna design includes a cut-out in the halved branch for a RF connector which is required for RF testing. CPW transmission lines were designed to match with the WBQD antenna (CPW impedance is 93Ω, center line width $w$ = 4 mm, gap width $g$ = 0.9 mm, and length $l$ = 44 mm) and were also used to test radio frequency conductivity of the printed graphene.

Graphene WBQD and graphene CPWs were screen-printed onto polyimide substrates (Kapton HN; DuPont; USA; 76 μm thickness) using a DEK Horizon 03i (DEK International, UK) semiautomatic screen printer following earlier published procedures [13]. In brief, graphene-flake inks were prepared by gelation of graphene/binder dispersions induced by mild heating with a graphene to binder ratio of 1:2. After solvent exchange the graphene ink was applied to the substrate by screen printing using a 45° angle polyurethane squeegee, at a printing speed of 50 mm/s followed by drying in air at 100 °C for 5 minutes. Subsequently, printed structures were thermally annealed at 350 °C for 30 minutes in air and, finally, compression rolled [1], [14] to reach the target DC sheet resistance of 4 Ω/□ at 10 μm layer thickness. For comparison, WBQDs and CPWs were also screen printed using a commercial silver ink onto electronic grade PET film (ST506; DuPont Teijin Films; thickness 125 μm). Printing was done with EKRA E2 semi-automatic screen and stencil printer (ASYS Group) in air at controlled temperature and relative humidity. The silver structures were dried at 130 °C for 30 minutes to reach a DC sheet resistance of 41 mΩ/□ at an approximate thickness of 50 μm. The printed WBQD antennas are shown in Fig. 1(b) and (c).

III. TESTING OF THE PRINTED SAMPLES

Before RF testing the WBQD and CPW structures were attached to a 1.5 mm thick Rohacell foam sheet for mechanical support as the foam has $\varepsilon_{r,R} \approx 1$ and, hence, a negligible effect on RF performance. Using an Agilent 8722ES vector network analyser (VNA) the S-parameters of the CPW test structures were measured from 1 GHz up to 20 GHz. The attenuation ($A$) of the CPW was calculated from the measured S-parameters following $A = (1-|S_{11}|^2)/|S_{21}|^2$ [3]. A comparison of the simulated and measured attenuation for the graphene CPWs and the measured attenuation for the silver CPW are shown in Fig. 2(a). Graphene CPW simulations with a sheet resistance of $R_s$ = 4 Ω/□ agree well with the measurements up to 5 GHz. At higher frequencies up to 13 GHz, agreement is good with simulations using $R_s$ = 5 Ω/□ and above 13 GHz, the results agree well with simulations using $R_s$ = 6 Ω/□ up to 19 GHz. The observed increase of sheet resistance with higher frequencies is probably caused by increased contact resistance of the highly porous graphene nanoflakes [1]. However, the results show that the sheet resistance of the printed graphene flakes is reasonable up to 20 GHz for developing antennas having moderate efficiencies.

The reflection coefficients ($S_{11}$) of the antennas were measured for the antenna frequency bandwidth from 1 GHz up to 5 GHz. A ferrite choke was used around the cable near the antenna feeding point to provide high impedance for the common-mode currents induced by the dipole antenna and this way reducing the cable effect in the measurements [15], [16]. The simulated and measured |$S_{11}$| of the antennas are presented in Fig. 2(b). It is seen that the measured |$S_{11}$| is below -8.5 dB from 1.7 GHz at least up to 5 GHz for both antennas, which indicates wideband performance. The results also show that over 85% of the input power is reaching the antenna and less than 15% is reflected back, so that the input matching is good and the antennas work as designed. Some deviation can be observed between the simulation and experimental results, which is due to common-mode currents that are not perfectly suppressed even when a ferrite choke is used. The graphene antenna has lower $|S_{11}|$ around 2 GHz, i.e., the input impedance is closer to 50 Ω than that of silver antenna. From 2.5 GHz to 5 GHz the results for silver and graphene antennas are very similar.

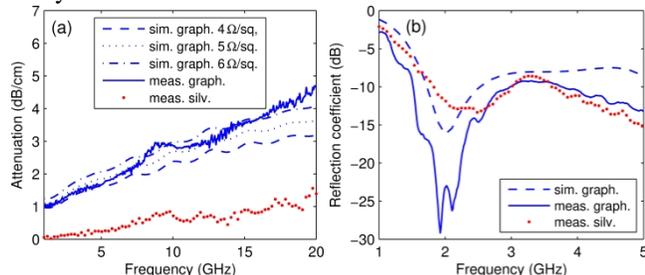

Fig. 2. (a) Simulated and measured attenuation for graphene CPW and measured attenuation for silver CPW. (b) Simulated and measured reflection coefficient for graphene WBQD antenna, and measured reflection coefficient for silver WBQD antenna.

The radiation patterns of the antennas were measured in an anechoic chamber. To minimise RF cable effects in the antenna measurements, an electro-optical link has been proposed [17]. An optical-to-RF transformer (ORT) with dimensions of 20 mm×23.5 mm×10 mm was also employed in this work. In the measurements, an optical signal is fed to the transmitting antenna under test (AUT) via an optical fiber to minimize the influence of RF cable radiation (Fig. 3(a)). The signal is transformed into an RF signal in the ORT block. The ORT output is directly connected to the AUT feeding RF connector. Here it should be noted that common-mode currents flow on the ORT block, which inevitably becomes a part of the antenna structure.

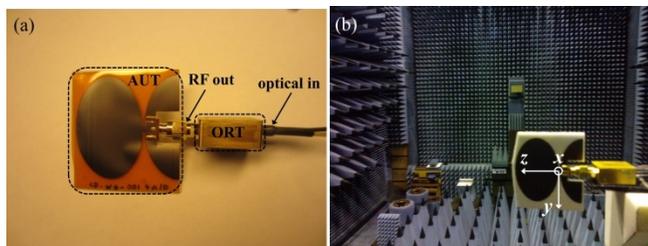

Fig. 3. (a) Image of graphene WBQD antenna connected to an optical-to-RF transformer. (b) Image of radiation pattern measurements in anechoic chamber.

The received signal was measured using a standard gain horn antenna and Agilent AG 8722E VNA analyser. The antenna is rotated by 360 deg. (-180 ≤ θ ≤ 180 deg.) in the horizontal plane with an antenna positioner by Scientific Atlanta and data is acquired in 2 degree steps. The antenna gain was determined using the gain comparison method with known standard gain reference antennas. Antenna efficiencies were determined from the measured directivity and the antenna gain. A photograph of the radiation pattern measurement setup is shown in Fig. 3(b).

The radiation patterns of the printed graphene-WBQD antenna were measured from 2 GHz up to 4.8 GHz. The ORT was included in simulations for precise comparison with the measurements. Simulated and measured co-polarised and cross-polarised gain patterns at 3 GHz are shown in Fig. 4(a) and (b) for two principal planes $\phi = 0$ degrees and $\phi = 90$ degrees, respectively. Typical dipole-like radiation patterns can be observed.

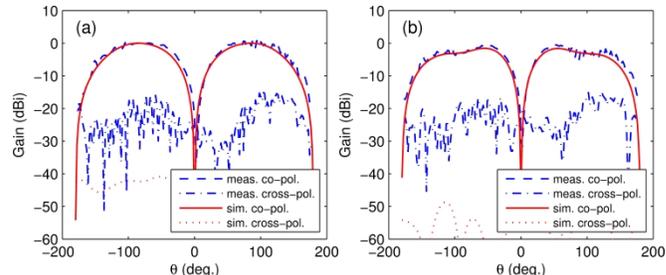

Fig. 4. Simulated and measured radiation patterns of the printed graphene WBQD antenna at 3 GHz in (a) $\phi = 0$ deg. plane and in (b) $\phi = 90$ deg. plane.

An excellent match between simulation and measurement is found in the co-polarised beam. The maximum directivity and gain are 3.3 dBi and 0.6 dBi resulting in an antenna efficiency of 56 ± 5%. The antenna has omni-directional coverage which is very suitable for mobile applications. The simulated cross-polarisation level is below -40 dBi while the measured level is approximately -17 dBi at maximum, which is due to a RF leakage by the ORT, increasing the gain in cross-polarisation. Despite the observed differences, the measured cross-polarisation level is sufficiently low that to warrant no effect on the overall antenna performance.

An important characteristics of the antenna is its directivity which describes the power density the antenna radiates over solid angle. Simulated and measured maximum directivity and maximum antenna gain are shown in Fig. 5(a) as a function of frequency. Overall, the simulations and measurements agree well. The curves are quite stable with frequency but some dips are observed at around 2–2.5 GHz. These dips are a measurement artefact and a result of a superposition of signals radiated by the antenna and the current flowing on the surface of the ORT metal housing. The effect was confirmed by simulations with and without ORT. Above 3 GHz, the surface currents on the ORT surface are very small and the radiation is excited only from the graphene-flakes printed antenna. It should be noted that in a real application an ORT would not be required. The antenna gain increases with frequency and has a maximum of 2.3 dBi at 4.8 GHz.

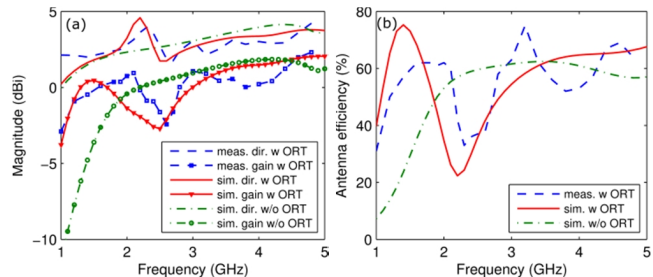

Fig. 5. Simulated and measured (a) maximum antenna gain and maximum directivity, and (b) antenna efficiency as a function of frequency of the printed graphene WBQD antenna.

Simulated and measured antenna efficiencies as a function of frequency are shown in Fig. 5(b). At frequencies from 2.8 GHz up to 4.8 GHz an antenna efficiency of 60% is measured. To the best of authors' knowledge, these are the highest gain and efficiency values published for graphene-flakes printed antennas to date. For comparison, the printed silver WBQD antenna was measured at 2 GHz and has a measured directivity and antenna gain of 2.7 dBi and 2.0 dBi, respectively, resulting in an antenna efficiency of 86%.

Even though the sheet resistance of printed graphene flakes is two orders of magnitude higher than the sheet resistance of printed silver, a total loss in an RF link budget (including transmitting and receiving sides) is only 3.1 dB higher than for a silver-printed antenna. Such a performance is adequate for usage of our developed graphene-flakes printed antenna in wireless communications applications with ranges from meters up to tens of meters.

Printed graphene layers were analyzed with respect to effects of repeated bending on their electrical resistance. For bending experiments a strip of material 11.3 cm long 0.9 cm wide was cut from the substrate and fixed with electrical contacts in the bending rig. During each of 32500 bending cycles (with a span of 40 mm and a bending radius of 42.5 mm per cycle) the resistance of the strip was measured. As seen in Fig. 6 we did not find any significant influence of repeated bending on the measured resistance.

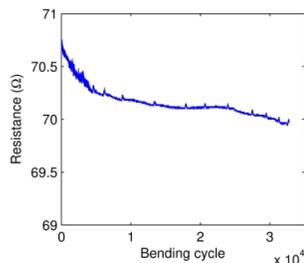

Fig. 6. Variation of resistance of graphene layer as a function of bending for 32500 cycles (with a span of 40 mm and a bending radius 42.5 mm per cycle).

The absolute resistance of the strip decreased from 70.73 Ω to 69.97 Ω, corresponding to about 1% gain in conductivity, indicating excellent fatigue resistance, at least for the described experimental conditions. For reference, screen printed silver on similar substrates have been shown to experience a gradual increase in resistance up to 10–20 % within a few hundred repetitions when bent to a radius of 6 mm [18].

A combination of a reasonable electrical performance with no degradation of conductive properties of the printed graphene-flakes layer even after 32500 bending cycles, renders the printed graphene flakes antennas suitable for low-cost wearable wireless communications devices such as health monitoring or smart clothing.

ACKNOWLEDGMENT

The authors would like to thank Mr. Ismo Huhtinen for the measurements and Mr. Alpo Ahonen for assembly work. Ms. Asta Pesonen is acknowledged for technical assistance with silver printing.

REFERENCES

[1] X. Huang, T. Leng T, X. Zhang, J. C. Chen, K. H. Chang, A. K. Geim, K. S. Novoselov, Z. Hu, "Binder-free highly conductive graphene laminate for low cost printed radio frequency applications," *Appl. Phys. Lett.* 106, 203105, pp.1–4, 2015.
[2] T. Leng, X. Huang, K. Chang, J. Chen, M. A. Abdalla, Z. Hu, "Graphene nanoflakes printed flexible meandered-line dipole antenna on paper substrate for low-cost RFID and sensing applications," *IEEE Antennas Wireless Propag. Lett.* vol. 15, pp. 1565–1568, 2016.
[3] X. Huang, T. Leng, M. Zhu, X. Zhang, J. Chen, K. Chang, M. Aqeeli, A. K. Geim, K. S. Novoselov, Z. Hu, "Highly flexible and conductive printed graphene for wireless wearable communications applications," *Sci. Rep.* 5:18298, pp. 1–7, 2015.
[4] K. Arapov, K. Jaakkola, V. Ermolov, G. Bex, E. Rubingh, S. Haque, H. Sandberg, R. Abbel, G. de With, H. Friedrich, "Graphene screen-printed radio-frequency identification devices on flexible substrates," *Phys. Status Solidi RRL*, pp. 1–7, 2016.
[5] M. Akbari, M. Waqas, A. Khan, M. Hasani, T. Björninen, L. Sydänheimo, L. Ukkonen, "Fabrication and characterization of graphene antenna for low-cost and environmentally friendly RFID tags," *IEEE Antennas Wireless Propag. Lett.* vol. 15, pp. 1569–1572, 2016.
[6] P. Kopyt, B. Salski, M. Olszewska-Placha, D. Janczak, M. Sloma, T. Kurkus, M. Jakubowska, W. Gwarek, "Graphene-based dipole antenna for a UHF RFID tag," *IEEE Trans. Antennas Propag.* vol. 64, pp. 2862–2868, 2016.
[7] J. Ryckaert, C. Desset, A. Fort, M. Badaroglu, V. D. Heyn, P. Wambacq, G. V. d. Plas, S. Donnay, B. V. Poucke, B. Gyselinckx, "Ultra-wide-Band transmitter for low-power wireless body area networks: design and evaluation," *IEEE Trans. Circuits Syst.* vol. 52, pp. 2515–2525, 2005.
[8] Q. H. Abbasi, A. Alomainy, Y. Hao, "Recent Development of Ultra wideband body-centric wireless communications," in *Proc. of the IEEE International Conference on Ultra-Wideband (ICUWB)*, Nanjing, China, 20–23 September 2010, pp. 1–4.
[9] N. P. Gupta, R. Maheshwari, M. Kumar, "Advancement in ultra wideband antennas for wearable applications," *IJSER*, vol. 4, pp. 341–348, 2013.
[10] X. Huang, T. Leng, K. H. Chang, J. C. Chen, K. S. Novoselov, Z. Hu, "Graphene radio frequency and microwave passive components for low cost wearable electronics," *2D Mater.*, vol. 3, 025021, pp. 1–8, 2016.
[11] H. G. Schantz, "Planar elliptical element ultra-wideband dipole antennas," in *IEEE AP-S Int. Symp. Dig.* vol.. 3, pp. 44–47, 2002.
[12] C. A. Balanis, *Antenna Theory: Analysis and Design* (John Wiley & Sons, Hoboken, 1997), p. 78.
[13] K. Arapov, E. Rubingh, R. Abbel, J. Laven, G. de With, H. Friedrich, "Conductive screen printing inks by gelation of graphene dispersions," *Adv. Funct. Mater.* 26, pp. 586–593, 2016.
[14] K. Arapov, G. Bex, R. Hendrix, E. Rubingh, R. Abbel, G. de With, H. Friedrich, "Conductivity enhancement of binder-based graphene inks by photonic annealing and subsequent compression rolling," *Adv. Eng. Mater.*, pp. 1–6, 2016.
[15] J. DeMarinis, "The antenna cable as a source of error in EMI measurements," in *Proc. of the IEEE International Symposium on Electromagnetic Compatibility (ISEMC)*, Seattle, Washington, 2–4 August 1988, pp. 9–14.
[16] T. W. Hertel, "Cable-current effects of miniature UWB antennas," in *Proc. of the IEEE Antennas and Propagation Society International Symposium*, Washington, DC, 3–8 July 2005, pp. 524–527.
[17] R. R. Lao, W. Liang, Y.-S. Chen, J. H. Tarng, "The use of electro-optical link to reduce the influence of RF cables in antenna measurement, in *Proc. of the IEEE International Symposium on Microwave, Antenna, Propagation and EMC Technologies for Wireless Communications (MAPE)*, Beijing, China, 8–12 August 2005, pp. 427–430.
[18] T. Happonen, T. Ritvonen, P. Korhonen, J. Häkkinen, T. Fabritius, "Bending reliability of printed conductors deposited on plastic foil with various silver pastes," *The International Journal of Advanced Manufacturing Technology*, 82, pp. 1663–1673, 2016.